\documentclass[twocolumn,prc,showpacs,superscriptaddress,preprintnumbers,nofootinbib]{revtex4}
\usepackage{amsmath,amssymb,longtable}
\usepackage{graphicx}
\usepackage{epsfig,color}
\usepackage{bm}
\usepackage{verbatim}
\usepackage{multirow}
\newcommand\nme[0]{\mathcal{M}}				  
\newcommand\vecs[1]{\boldsymbol{#1}}     		  

\begin{document}

\title{Nuclear matrix elements for Majoron emitting double-$\beta$ decay}
\author{J. Kotila}\email[]{jenni.kotila@jyu.fi}
\affiliation{Finnish Institute for Educational Research, University of Jyv\"askyl\"a, P.O. Box 35, 40014 Jyv\"askyl\"a, Finland}
\affiliation{Center for Theoretical Physics, Sloane Physics Laboratory
Yale University, New Haven, Connecticut 06520-8120, USA}

\author{ F.\ Iachello}\email[]{francesco.iachello@yale.edu}
\affiliation{Center for Theoretical Physics, Sloane Physics Laboratory
Yale University, New Haven, Connecticut 06520-8120, USA}

\begin{abstract}
A complete calculation of the nuclear matrix elements (NME) for Majoron
emitting neutrinoless double beta decay within the framework of IBM-2 for
spectral indices $n=1,3,7$ is presented. By combining the results of this
calculation with previously calculated phase space factors (PSF) we give
predictions for expected half-lives. By comparing with experimental limits
on the half-lives we set limits on the coupling constants $\left\langle
g_{ee}^{M}\right\rangle $ of all proposed Majoron-emitting models.
\end{abstract}

\maketitle

\section{Introduction}

In recent years, increased accuracy has been achieved in the measurement of
double beta decay (DBD) with the emission of two neutrinos, $2\nu \beta
\beta $ decay, especially in the measurement of the summed electron spectra.
High statistics experiments have been reported by GERDA ($^{76}$Ge) \cite{GERDA}, NEMO3 ($^{100}$Mo) \cite{NEMO3}, CUORE ($^{130}$Te) \cite{CUORE},
EXO ($^{136}$Xe) \cite{EXO} and KamLAND-Zen ($^{136}$Xe) \cite{KamLAND}.
High statistics experiments have provided information on the mechanism of
DBD in $2\nu \beta \beta $ decay, in particular the question of single state
dominance (SSD) versus high state dominance (HSD), CUPID-0 ($^{82}$Se) \cite{CUPID-0} and CUPID-Mo ($^{100}$Mo) \cite{CUPID-Mo}. With the degree of
accuracy reached in the latest experiments, one can also test non-standard
mechanisms of DBD and set stringent limits on them \cite{dep2020}.

\begin{figure}[htbp]
\begin{center}
\includegraphics[width=0.48\textwidth]{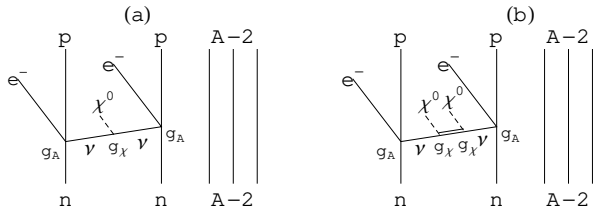}
\caption{Schematic representation of neutrinoless double-beta decay
accompanied by the emission of one or two Majoron. Adapted from \cite{kotila-maj}.
}
\label{fig1}
\end{center}
\end{figure}

One of the non-standard mechanisms is that occurring with the emission of
additional bosons called Majorons. Majorons were introduced years ago \cite{Chika,gelmini} as massless Nambu-Goldstone bosons arising from
global $B-L$ (baryon number minus lepton number symmetry) broken
spontaneously in the low-energy regime. These bosons couple to the Majorana
neutrinos and give rise to neutrinoless double beta decay, accompanied by
Majoron emission $0\nu \beta \beta M$ \cite{georgi}, as schematically shown
in Fig.\ref{fig1}(a). Although these older models are disfavored by precise
measurements of the width of the Z boson decay to invisible channels \cite%
{CERN}, several other models of $0\nu \beta \beta M$ decay have been
proposed in which one or two Majorons, denoted by $\chi _{0}$, are emitted
(see Fig.\ref{fig1}):%
\begin{eqnarray}
(A,Z) &\rightarrow &(A,Z+2)+2e^{-}+\chi _{0}  \nonumber \\
(A,Z) &\rightarrow &(A,Z+2)+2e^{-}+2\chi _{0}.
\end{eqnarray}%
Table \ref{tableI} lists some of the models proposed to describe these decays \cite{bamert, carone, burgess, mohapatra}. The different
models are distinguished by the nature of the emitted Majoron(s), i.e.
whether it is a Nambu-Goldstone boson or not (NG), the leptonic charge of
the emitted Majoron (L), and the spectral index of the model, $n$.

\medskip

\begin{table}
\begin{tabular}{cccccc}
\hline
\text{Model} & \text{Decay mode} & \text{NG boson} & L & n & \text{NME} \\ 
\hline 
IB & $0\nu \beta \beta \chi _{0}$ & \text{No} & 0 & 1 & $M_{1}$ \\ 
IC & $0\nu \beta \beta \chi _{0}$ & \text{Yes} & 0 & 1 & $M_{1}$ \\ 
ID & $0\nu \beta \beta \chi _{0}\chi _{0}$ & \text{No} & 0 & 3 & $M_{3}$ \\ 
IE & $0\nu \beta \beta \chi _{0}\chi _{0}$ & \text{Yes} & 0 & 3 & $M_{3}$ \\ 
IIB & $0\nu \beta \beta \chi _{0}$ & \text{No} & -2 & 1 & $M_{1}$ \\ 
IIC & $0\nu \beta \beta \chi _{0}$ & \text{Yes} & -2 & 3 & $M_{2}$ \\ 
IID & $0\nu \beta \beta \chi _{0}\chi _{0}$ & \text{No} & -1 & 3 & $M_{3}$ \\ 
IIE & $0\nu \beta \beta \chi _{0}\chi _{0}$ & \text{Yes} & -1 & 7 & $M_{3}$ \\ 
IIF & $0\nu \beta \beta \chi _{0}$ & \text{Gauge boson} & -2 & 3 & $M_{2}$ \\ 
"Bulk" & $0\nu \beta \beta \chi _{0}$ & \text{Bulk field} & 0 & 2 & -\\
\hline
\end{tabular}%
\caption{Different Majoron emitting models \cite{bamert, carone, burgess, mohapatra}. The third, fourth, and
fifth columns indicate whether the Majoron is a Nambu-Golstone boson or not,
its leptonic charge $L$, and the model's spectral index, $n$. The sixth
column indicates the nuclear matrix elements of Sect. II appropriate for
each model.
}
\label{tableI}
\end{table}
\medskip

The half-life for all these models can be written as 
\begin{equation}
\label{hl}
\left[ \tau _{1/2}^{0\nu M}\right] ^{-1}=G_{m\chi _{0}n}^{(0)}\left\vert
\left\langle g_{\chi _{ee}^{M}}\right\rangle \right\vert ^{2m}\left\vert
M_{0\nu M}^{(m,n)}\right\vert ^{2}
\end{equation}%
where $G_{m\chi _{0}n}^{(0)}$ is a phase space factor (PSF), $\left\langle
g_{\chi _{ee}^{M}}\right\rangle $ the effective coupling constant of the
Majoron to the neutrino, $m=1,2$ for the emission of one or two Majorons,
respectively, and $M_{0\nu M}^{(m,n)}$ the nuclear matrix element (NME).

\section{Phase space factors}

\begin{figure}[htbp]
\begin{center}
\includegraphics[width=18pc]{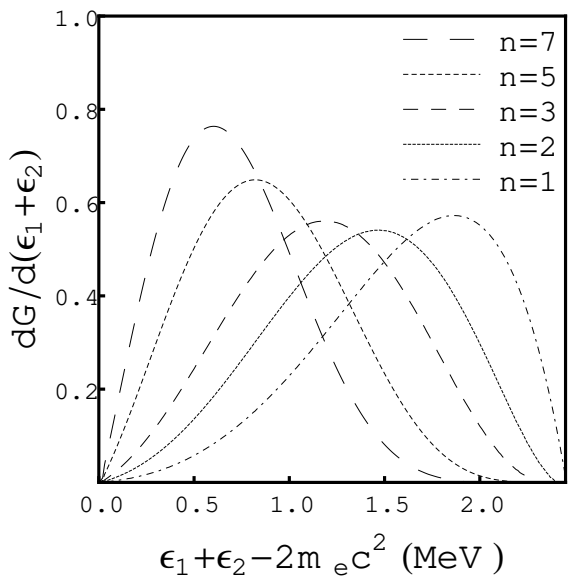}
\caption{Summed electron spectra for the $n=1,2,3$ and $7$, as well as for
the $2\nu\beta\beta$ ($n=5$) decays of $^{136}$Xe.}
\label{fig2}
\end{center}
\end{figure}

\begin{figure}[htbp]
\begin{center}
\includegraphics[width=18pc]{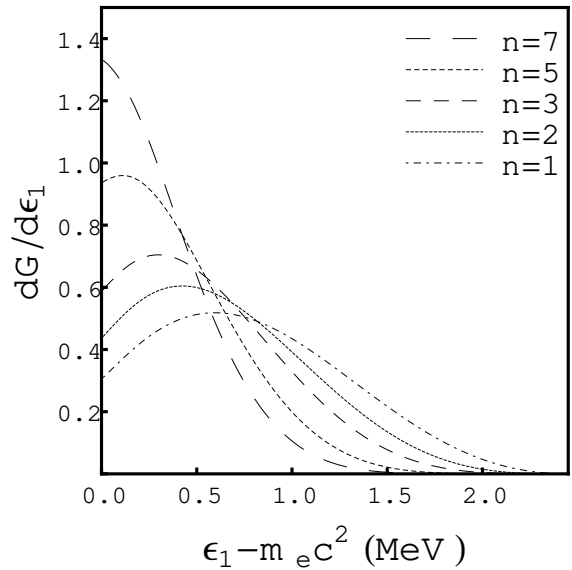}
\caption{Single electron spectra for the $n=1,2,3$ and $7$, as well as for
the $2\nu\beta\beta$ ($n=5$) decays of $^{136}$Xe.
}
\label{fig3}
\end{center}
\end{figure}

\begin{figure}[htbp]
\begin{center}
\includegraphics[width=16pc]{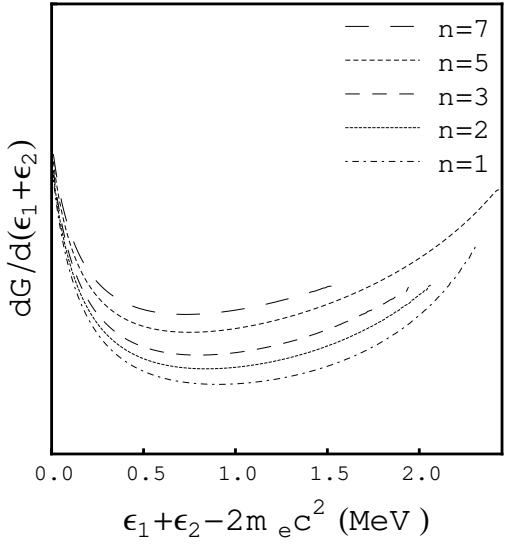}
\caption{ Angular correlation between the two emitted electrons for the $n=1,2,3$ and $7$, as well as for the $2\nu\beta\beta$ ($n=5$) decays of $^{136}$Xe. The calculation for $n=1,2,3,7$ stops at the point where the single electron spectrum goes to zero. Beyond that point it becomes unstable as $G^{(0)}$ goes to zero faster than $G^{(1)}$. For $2\nu\beta\beta$ ($n=5$) this is avoided by taking into account the individual energies of the neutrinos, $\omega_1$ and $\omega_2$, as in Eq.(22) of \cite{kotila}, instead of the Majoron energy $q$ as in Eq.(5) of \cite{kotila-maj}.
}
\label{fig4}
\end{center}
\end{figure}

In a previous article \cite{kotila-maj} we have calculated the PSF and from
these the single electron spectrum, the summed electron spectrum and the
angular correlation between the two electrons. Particularly interesting are
the summed electron spectra whose shape depends crucially on the spectral
index $n$. In Fig.\ref{fig2}, the summed electron spectra for $n=1,n=3$ and $n=7$,
obtained from \cite{kotila-maj} by normalizing the spectra so that the area
covered by each of them is the same, are plotted as a function of $%
\varepsilon _{1}+\varepsilon _{2}-2m_{e}c^{2}$. In this figure, also the
summed electron spectrum for $2\nu \beta \beta $ decay \cite{kotila} is
shown again with area normalized to 1. This spectrum has a spectral index $%
n=5$. The summed electron spectrum of the "bulk" model $n=2$ is also shown
in Fig.\ref{fig2}. 
Exact Dirac wave functions, nuclear finite size and electron
screening are included in this calculation as discussed in \cite{kotila}.
Previous calculations \cite{doi, tomoda, suhonen, hirsch} make use of Fermi functions which are an approximation to the
relativistic Dirac wave functions. For comparison between the values
reported in \cite{hirsch} and our values \cite{kotila-maj} we note that our
PSF are divided by a factor of $g_{A}^{4}=2.593$ since we include this
factor in the NME. We have estimated the error in using the old calculation
of the PSFs  \cite{doi, tomoda, suhonen, hirsch} instead of the new \cite{kotila-maj}, $\left( G_{m\chi
_{0}n}^{(0)old}-G_{m\chi _{0}n}^{(0)new}\right) /G_{m\chi _{0}n}^{(0)new}$,
to be 6\% in $^{76}$Ge and 28\% in $^{136}$Xe. The reason why the error is
larger in $^{136}$Xe($Z=54$) than in $^{76}$Ge($Z=32$) is the neglect in the
old calculation of relativistic effects and electron screening which
increase as a large power of $Z$. While in $^{76}$Ge and $^{82}$Se the use
of the old calculation may still be reasonable, it is definitely not so in
the other nuclei of current interest $^{100}$Mo, $^{130}$Te, $^{136}$Xe and $^{150}$Nd. Although experimentally not easily accessible, we also plot in
Fig.\ref{fig3} the single electron spectra with area normalized to 1, and in Fig.\ref{fig4}
the angular correlation between the two electrons, for $n=1,n=2,n=3,n=7$ \
and $2\nu \beta \beta $ ($n=5$) as a function of $\varepsilon _{1}-m_{e}c^{2}
$, both of which have been measured by the NEMO3 collaboration in $^{130}$Te 
\cite{NEMO3-1}. In this article we present a calculation of the nuclear
matrix elements $M_{0\nu M}^{(m,n)}$.

\section{Nuclear matrix elements}

Nuclear matrix elements for Majoron emitting DBD were derived in a seminal
paper by Hirsch \textit{et al.} \cite{hirsch}. These authors derived an
explicit form for the nuclear matrix elements of all the models of Table \ref{tableI},
except the "bulk" model. We have converted the form of \cite{hirsch} to our
notation, added some higher order terms not included in the original form
and calculated the corresponding matrix elements within the framework of the
microscopic interacting boson model IBM-2 \cite{barea1, barea2} with
isospin restoration \cite{barea3}. Explicitly, we introduce the matrix
elements%
\begin{eqnarray}
\label{FGTT}
\mathcal{M}_{F} &=&\left\langle f\left\Vert v_{m}\right\Vert i\right\rangle 
\nonumber \\
\mathcal{M}_{GT} &=&\left\langle f\left\Vert v_{m}\vecs{\sigma}_1\cdot\vecs{\sigma}_2 \right\Vert i\right\rangle  \nonumber \\
\mathcal{M}_{T} &=&\left\langle f\left\Vert v_{m}S_{12}\right\Vert
i\right\rangle  \nonumber \\
\mathcal{M}_{GTR} &=&\left\langle f\left\Vert v_{R}\vecs{\sigma}_1\cdot\vecs{\sigma}_2 \right\Vert i\right\rangle \\
\mathcal{M}_{TR} &=&\left\langle f\left\Vert v_{R}S_{12}\right\Vert
i\right\rangle  \nonumber \\
\mathcal{M}_{F\omega ^{2}} &=&\left\langle f\left\Vert v_{\omega
^{2}}\right\Vert i\right\rangle  \nonumber \\
\mathcal{M}_{GT\omega ^{2}} &=&\left\langle f\left\Vert v_{\omega ^{2}}%
\vecs{\sigma}_1\cdot\vecs{\sigma}_2 \right\Vert i\right\rangle 
\nonumber \\
\mathcal{M}_{T\omega ^{2}} &=&\left\langle f\left\Vert v_{\omega
^{2}}S_{12}\right\Vert i\right\rangle  \nonumber
\end{eqnarray}%
where the isospin operators $\tau _{1}^{+}\tau _{2}^{+}$ have been dropped
for simplicity. These matrix elements are the same as in \cite{hirsch} with
the addition of the tensor matrix elements.

The neutrino potentials needed for the calculation of these matrix elemnts,
when converted to the notation used in IBM-2 \cite{barea1,barea2,barea3} are%
\begin{eqnarray}
\label{potentials}
v_{m} &=&\frac{2}{\pi }\frac{1}{q(q+\tilde{A})}  \nonumber \\
v_{R} &=&\frac{2}{\pi }\frac{1}{Rm_{p}}\frac{q+\frac{\tilde{A}}{2}}{q(q+%
\tilde{A})^{2}} \\
v_{\omega ^{2}} &=&\frac{2}{\pi }m_{e}^{2}\frac{q^{2}+\frac{9}{8}q\tilde{A}+%
\frac{3}{8}\tilde{A}^{2}}{q^{3}(q+\tilde{A})^{3}}  \nonumber
\end{eqnarray}%
with $R=1.2A^{1/3}$fm, $m_{p}=938$ MeV$=4.76$ fm$^{-1}$, $m_{e}=0.511$ MeV$%
=0.00259$ fm$^{-1}$. $\tilde{A}$ is the closure energy that we take as in 
\cite{barea1,barea2,barea3} $\tilde{A}=1.12A^{1/2}$ MeV,
where $A$ denotes the mass number. We note that the last term in $v_{\omega
^{2}}$ diverges at the origin as $q^{-3}$. We regularize this term by
multiplying it by $q/(q+\tilde{A})$, that is%
\begin{equation}
v_{\omega ^{2}}=\frac{2}{\pi }m_{e}^{2}\frac{q^{2}+\frac{9}{8}q\tilde{A}+%
\frac{3}{8}\tilde{A}^{2}\frac{q}{q+\tilde{A}}}{q^{3}(q+\tilde{A})^{3}}.
\end{equation}%
From the neutrino potentials we construct the quantities%
\begin{equation}
h(q)=v(q)\tilde{h}(q)
\end{equation}%
where $\tilde{h}_{F,GT,T}(q)=\tilde{h}_{F\omega ^{2},GT\omega ^{2},T\omega
^{2}}$ are given in Table II of \cite{barea2} which includes the form
factors and higher order corrections and 
\begin{equation}
\tilde{h}_{R}(q)=\frac{1}{(1+q^{2}/m_{V}^{2})^{2}}\frac{1}{%
(1+q^{2}/m_{A}^{2})^{2}}
\end{equation}%
which includes the form factors with $m_{V}=0.84$ GeV and $m_{A}=1.09$ GeV
as in \cite{barea2,barea3}.

The matrix elements for the three classes of Majoron models are%
\begin{align}
\label{nmes}
M_{1}&=g_{A}^{2}\mathcal{M}_{1}=g_{A}^{2}\left[ -\left( \frac{g_{V}^{2}}{%
g_{A}^{2}}\right) \mathcal{M}_{F}+\mathcal{M}_{GT}-\mathcal{M}_{T}\right] 
 \\
M_{2}&=g_{A}^{2}\mathcal{M}_{2}=g_{A}^{2}\left[ \left( \frac{g_{V}}{g_{A}}%
\right) \frac{f_{W}}{3}\mathcal{M}_{GTR}-\left( \frac{g_{V}}{g_{A}}\right) 
\frac{f_{W}}{6}\mathcal{M}_{TR}\right] \nonumber\\
M_{3}&=g_{A}^{2}\mathcal{M}_{3}=g_{A}^{2}\left[ -\left( \frac{g_{V}^{2}}{%
g_{A}^{2}}\right) \mathcal{M}_{F\omega ^{2}}+\mathcal{M}_{GT\omega ^{2}}-%
\mathcal{M}_{T\omega ^{2}}\right]  \nonumber 
\end{align}%
where we have used the overall sign convention as in \cite{simkovic} and in
our previous papers \cite{barea1,barea2,barea3}. In Eq.(\ref{nmes}), $f_{W}=1+\kappa _{\beta }=4.70$, where $\kappa _{\beta }$ is the isovector
magnetic moment of the nucleon. In the calculation of the matrix elements in
Eq.(\ref{nmes}) also short range correlations are included as in \cite{barea1,barea2,barea3}. Our results are shown in Table \ref{tab:majoronnme}. The nuclear
matrix elements $M_{1},M_{2},M_{3}$ are associated with Majoron emitting
models of $0\nu \beta \beta M$ decays as in last column of Table \ref{tableI}.

\medskip

\begin{table*}[t!]
	\centering
	\begin{tabular}{@{\extracolsep{4pt}}r|cccc|ccc|cccc@{}}
		\hline
		Isotope 		& \multirow{2}{*}{$\nme_F$} & \multirow{2}{*}{$\nme_{GT}$} &\multirow{2}{*}{ $\nme_T$} & \multirow{2}{*}{$\nme_1$} &\multirow{2}{*}{ $\nme_{GTR}$} & \multirow{2}{*}{$\nme_{TR}$}  &\multirow{2}{*}{ $\nme_2$} &$\nme_{F\omega^2}$ & $\nme_{GT\omega^2}$ & $\nme_{T\omega^2}$ & $\nme_3$\\ 
					&		&			&			&		&				&			&			&$\times10^{3}$				&$\times10^{3}$					&$\times10^{3}$	&$\times10^{3}$\\\hline
		${}^{76}$Ge	& $-0.780$	& $5.582$	& $-0.281$	& $6.642$		&$0.225$	&$-0.037$	&$0.381$		&$-0.017$		&$2.530$	&$-0.009$	&	$2.556$	\\
		${}^{82}$Se	& $-0.667$	& $4.521$	& $-0.270$	& $5.458$		&$0.178$	&$-0.034$	&$0.305$		&$-0.014$		&$1.967$	&$-0.009$	&	$1.993$\\
		${}^{96}$Zr	& $-0.361$	& $3.954$	& $0.250$		& $4.065$		&$0.147$	&$0.031$	&$0.205$		&$-0.006$		&$1.672$	&$0.009$		&$1.668$\\
		${}^{100}$Mo	& $-0.511$	& $5.075$	& $0.318$		& $5.268$		&$0.187$	&$0.038$	&$0.263$		&$-0.008$		&$1.904$	&$0.011$		&$1.901$\\
		${}^{110}$Pd	 &$-0.425$	& $4.024$	& $0.243$		& $4.206$		&$0.144$	&$0.030$	&$0.203$		&$-0.006$		&$1.411$	&$0.009$		&$1.409$\\
		${}^{116}$Cd	& $-0.335$	& $2.888$	& $0.118$		& $3.105$		&$0.102$	&$0.019$	&$0.144$		&$-0.005$		&$0.945$	&$0.006$		&$0.945$\\
		${}^{124}$Sn	& $-0.572$	& $3.099$	& $-0.118$	& $3.789$		&$0.104$	&$-0.017$	&$0.177$		&$-0.013$		&$1.161$	&$-0.005$		&$1.179$\\
		${}^{128}$Te	& $-0.718$	& $3.965$	& $-0.115$	& $4.798$		&$0.132$	&$-0.020$	&$0.223$		&$-0.016$		&$1.505$	&$-0.006$		&$1.527$\\
		${}^{130}$Te	& $-0.651$	& $3.586$	& $-0.159$	& $4.396$		&$0.118$	&$-0.018$	&$0.199$		&$-0.014$		&$1.291$	&$-0.005$		&$1.311$\\
		${}^{134}$Xe	& $-0.686$	& $3.862$	& $-0.121$	& $4.669$		&$0.126$	&$-0.018$	&$0.212$		&$-0.015$		&$1.456$	&$-0.006$		&$1.477$\\
		${}^{136}$Xe	& $-0.522$	& $2.958$	& $-0.123$	& $3.603$		&$0.096$	&$-0.013$	&$0.160$		&$-0.012$		&$1.161$	&$-0.004$		&$1.112$\\
		${}^{148}$Nd	& $-0.362$	& $2.283$	& $0.125$		& $2.521$		&$0.074$	&$0.012$	&$0.107$		&$-0.006$		&$0.648$	&$0.004$		&$0.650$\\
		${}^{150}$Nd	& $-0.507$	& $3.371$	& $0.119$		& $3.759$		&$0.110$	&$0.017$	&$0.159$		&$-0.008$		&$0.836$	&$0.005$		&$0.839$\\
		${}^{154}$Sm	& $-0.340$	& $2.710$	& $0.122$		& $2.928$		&$0.086$	&$0.015$	&$0.122$		&$-0.006$		&$0.858$	&$0.005$		&$0.859$\\
		${}^{160}$Gd	& $-0.415$	& $3.838$	& $0.250$		&$4.002$		&$0.120$	&$0.023$	&$0.170$		&$-0.006$		&$1.261$	&$0.008$		&$1.260$\\
		${}^{198}$Pt	& $-0.329$	& $2.021$	& $0.119$		&$2.230$		&$0.061$	&$0.009$	&$0.089$		&$-0.005$		&$0.393$	&$0.003$		&$0.395$\\
		${}^{232}$Th	& $-0.444$	& $3.757$	& $0.251$		& $3.950$		&$0.104$	&$0.019$	&$0.148$		&$-0.006$		&$0.930$	&$0.007$		&$0.930$\\
		${}^{238}$U	& $-0.525$	& $4.470$	& $0.244$		&$4.751$		&$0.122$	&$0.022$	&$0.174$		&$-0.007$		&$1.118$	&$0.008$		&$1.118$\\
		\hline
	\end{tabular}
	\caption{Majoron emitting DBD NMEs $\mathcal{M}_{i}(i=1,2,3)$ calculated in
this work using the quenched value $g_A = 1.0$ and the convention that $\nme_i > 0$. }
	\label{tab:majoronnme}
\end{table*}

\medskip

\subsection{Sensitivity to parameter changes, model assumptions and operator
assumptions}

The matrix element $\mathcal{M}_{1}$ for index $n=1$ is identical to the
matrix element of ordinary $0\nu \beta \beta $ decay without Majoron
emission. \ The sensitivity of IBM-2 calculations to parameter changes,
model assumptions and operator assumption for this NME was discussed in
great detail in \cite{barea2,barea3}. Our error estimate for $M_{1}$
is therefore 16\% for all nuclei as in \cite{barea3}.

For the matrix element $\mathcal{M}_{2}$ we have an additional error coming
from the neglect of higher order terms of the type%
\begin{equation}
\begin{split}
\frac{(\mathbf{Q\cdot \vecs{\sigma}}_1)(\mathbf{q\cdot \vecs{\sigma}}_2)}{4m_{p}^{2}}%
& \simeq \frac{(\mathbf{q\cdot \vecs{\sigma}}_1)(\mathbf{q\cdot \vecs{\sigma}}_2)}{%
4m_{p}^{2}} \\
&=\frac{q^{2}}{4m_{p}^{2}}\left[ \frac{1}{3}\vecs{\sigma}_1\cdot\vecs{\sigma}_2 +\frac{1}{3}S_{12}\right] ,
\end{split}%
\end{equation}
where $\mathbf{Q}$ is the total momentum and $\mathbf{q}$ the relative
momentum of the nucleons and we have assumed $\mathbf{Q\simeq q}$ \cite%
{tomoda}. We estimate the neglected contribution of these higher order terms
to be about 4\% giving a total estimated error of 20\% for $\mathcal{M}_{2}$.

The matrix element $\mathcal{M}_{3}$ depends strongly on the closure energy $%
\tilde{A}$ as given in Eq.(\ref{potentials}). In the present calculation we have assumed
the standard choice $\tilde{A}=1.12A^{1/2}$ MeV. We have investigated
variations of $\tilde{A}$ around the standard values and estimate an
additional error in the calculation of $\mathcal{M}_{3}$ of $\sim $10\%
bringing the total estimated error to 30\%. An estimate of the sensitivity
of $\mathcal{M}_{3}$ to the closure energy was also given in \cite{hirsch}.
In this reference also a discussion of the sensitivity to model assumptions
of Majoron emitting DBD was given.

\section{Limits on the coupling constants}

From the PSF of \cite{kotila-maj}, the NME of this article, and experimental
limits on half-lives for each type of Majoron model, one can derive limits
on the coupling constants $g_{\chi _{ee}^{M}}$. These limits depend on the
value of the coupling constant $g_{A}$. This coupling constant is
renormalized in nuclei by many-body effects. Three possible values are \cite%
{delloro}: (i) the free value, $g_{A}=1.269$, (ii) the quark value, $%
g_{A}=1.0$, and (iii) the value extracted from $2\nu \beta \beta $ decay,
which, in IBM-2 can be parametrized as $g_{A,eff}^{IBM-2}=1.269A^{-0.18}$.
In order to allow for different values of $g_{A}$, we rewrite Eq.(\ref{hl}) as%
\begin{equation}
\label{half-life}
\left[ \tau _{1/2}^{0\nu M}\right] ^{-1}=G_{m\chi _{0}n}^{(0)}\left\vert
\left\langle g_{\chi _{ee}^{M}}\right\rangle \right\vert
^{2m}g_{A}^{4}\left\vert \mathcal{M}_{0\nu M}^{(m,n)}\right\vert ^{2}
\end{equation}%
where $\mathcal{M}_{0\nu M}^{(m,n)}$ are the NME given in Table \ref{tab:majoronnme}. In
extracting limits on $g_{\chi _{ee}^{M}}$ we use in this article $g_{A}=1$.
From Eq.(\ref{half-life}) it is straightforward to obtain limits for other values of $%
g_{A} $.

Limits on half-lives for Majoron emitting models have been reported by
several groups \cite{CUORE-1,NEMO3-1,KamLAND-1,EXO-1,GERDA-1}. In Table \ref{tableIII} we provide our limits on the coupling
constants $g_{\chi _{ee}^{M}}$.

\medskip

\begin{table*}[t!]
	\centering
	\begin{tabular}{@{\extracolsep{4pt}}ccccccc@{}}
		\hline
Decay mode &Spectral Index &Model Type &$ \mathcal{M}$ &$ G_{m\chi _{0}n}^{(0)}[10^{-18}yr]$ &$ \tau _{1/2}[yr]$ &$ \left\vert <g_{\chi _{ee}^{M}}>\right\vert $\\ 
\hline
$^{76}$\text{Ge \cite{GERDA-1}} &  &  &  &  &  &  \\ 
$0\nu \beta \beta \chi _{0}$ & 1 & \text{IB,IC,IIB} & 6.64 			& 44.2 &$ >4.2\times10^{23}$ &$ <3.5\times 10^{-5}$ \\ 
$0\nu \beta \beta \chi _{0}\chi _{0}$ & 3 & \text{ID,IE,IID} & 0.0026 	& 0.22&$ >0.8\times 10^{23}$ &$ <1.7 $ \\ 
$0\nu \beta \beta \chi _{0}$ & 3 & \text{IIC,IIF} & 0.381 			& 0.073 & $>0.8\times10^{23}$ &$<0.34 \times 10^{-1}$\\ 
$0\nu \beta \beta \chi _{0}\chi _{0}$ & 7 & \text{IIE} & 0.0026 		& 0.420 & $>0.3\times 10^{23}$ &$ <1.9 $\\ 
$0\nu \beta \beta \chi _{0}$ & 2 & \text{Bulk} & - &  				&$ >1.8\times 10^{23}$ & \\
\hline
$^{130}$\text{Te \cite{CUORE-1}} &  &  &  &  &  &  \\ 
$0\nu \beta \beta \chi _{0}$ & 1 & \text{IB,IC,IIB} & 4.40 			& 413 & $>2.2\times10^{21}$ & $<2.4\times 10^{-4} $\\ 
$0\nu \beta \beta \chi _{0}\chi _{0}$ & 3 & \text{ID,IE,IID} & 0.0013 	& 3.21 & $>0.9\times 10^{21}$ & $<3.8$ \\ 
$0\nu \beta \beta \chi _{0}$ & 3 & \text{IIC,IIF} & 0.199 			& 1.51 & $>2.2\times10^{21}$ & $<0.87\times 10^{-1}$ \\ 
$0\nu \beta \beta \chi _{0}\chi _{0}$ & 7 & \text{IIE} & 0.0013 		& 14.4 & $>0.9\times 10^{21}$ & $<2.6$ \\ 
$0\nu \beta \beta \chi _{0}$ & 2 & \text{Bulk} & - &  &$ >2.2\times 10^{21}$ & \\ 
\hline
$^{130}$\text{Te \cite{NEMO3-1}} &  &  &  &  &  &  \\ 
$0\nu \beta \beta \chi _{0}$ & 1 & \text{IB,IC,IIB} & 4.40 & 413 & $>1.6\times10^{22}$ & $<8.8\times 10^{-5}$\\%
\hline
$^{136}$\text{Xe \cite{EXO-1}} &  &  &  &  &  &  \\ 
$0\nu \beta \beta \chi _{0}$ & 1 & \text{IB,IC,IIB} & 3.60 & 409 & $>1.2\times10^{24}$ & $<1.3\times 10^{-5} $\\ 
$0\nu \beta \chi _{0}\chi _{0}$ & 3 & \text{ID,IE,IID} & 0.0011 & 3.05 & $>2.7\times 10^{22}$ & $<1.8$ \\ 
$0\nu \beta \beta \chi _{0}$ & 3 & \text{IIC,IIF} & 0.160 & 1.47 & $>2.7\times10^{22}$ & $<0.31\times 10^{-1}$ \\ 
$0\nu \beta \beta \chi _{0}\chi _{0}$ & 7 & \text{IIE} & 0.0011 & 12.5 & $>6.1\times 10^{21}$ & $<1.8$ \\ 
$0\nu \beta \beta \chi _{0}$ & 2 & \text{Bulk} & - & - & $>2.5\times 10^{23}$ & \\
\hline
$^{136}$\text{Xe \cite{KamLAND-1}} &  &  &  &  &  &  \\ 
$0\nu \beta \beta \chi _{0}$ & 1 & \text{IB,IC,IIB} & 3.60 & 409 & $>2.6\times10^{24}$ & $<8.5\times 10^{-6}$ \\ 
$0\nu \beta \beta \chi _{0}\chi _{0}$ & 3 & \text{ID,IE,IID} & 0.0011 & 3.05 & $>4.5\times 10^{24}$ & $<0.49$ \\ 
$0\nu \beta \beta \chi _{0}$ & 3 & \text{IIC,IIF} & 0.160 & 1.47 & $>4.5\times10^{24}$ & $<0.24\times 10^{-2}$ \\ 
$0\nu \beta \beta \chi _{0}\chi _{0}$ & 7 & \text{IIE} & 0.0011 & 12.5 & $>1.1\times 10^{22}$ & $<1.6$ \\ 
$0\nu \beta \beta \chi _{0}$ & 2 & \text{Bulk} & - & - &$ >1.0\times 10^{24}$ & -\\
\hline
\end{tabular}
\caption{Limits on the Majoron-neutrino coupling constants $\left\langle
g_{\chi _{ee}^{M}}\right\rangle $ for $g_{A}=1$. PSF from \cite{kotila-maj}.
NME from this paper.
}
\label{tableIII}
\end{table*}

\medskip

The most stringent limits come from the KamLAND-Zen collaboration \cite%
{KamLAND-1} and from the EXO collaboration \cite{EXO-1}. The reason why one
obtains such small limits for Majoron emitting models with index $n=1$ was
discussed in \cite{hirsch}. The larger limits of $g_{\chi _{ee}^{M}}$ for
Majoron emitting models with index $n=3$ and $n=7$ are due to the smaller
values of the PSF for these indices.

\section{Conclusions}

We have presented here a complete calculation of NME for Majoron emitting
neutrinoless double beta decay within the framework of the Interacting Boson
Model IBM-2. Our results when combined with the phase space factors of 
\cite{kotila-maj} provide up-to-date predictions for life-times, single
electron spectra, summed electron spectra and angular distributions for
Majoron emitting neutrinoless double beta decay which can be used in the
analysis of recent high statistics experiments \cite{GERDA,NEMO3,CUORE,EXO,KamLAND,CUPID-0,CUPID-Mo}.

\section{Acknowledgements}

This work was supported in part by the Academy of Finland Grant Nos. 314733,
320062.

\bigskip


\begin{thebibliography}{99}
\bibitem{GERDA}  M. Agostini \textit{et al.} (The GERDA Collaboration), Nature 
\textbf{544}, 47 (2017).

\bibitem{NEMO3}  R. Arnold \textit{et al.} (The NEMO3 Collaboration), Phys.
Rev. D \textbf{92}, 072011 (2015).

\bibitem{CUORE}  K. Alfonso \textit{et al.} (The CUORE Collaboration), Phys.
Rev. Lett. \textbf{115}, 102502 (2015).

\bibitem{EXO}  N. Ackerman \textit{et al.} (The EXO\ Collaboration), Phys.
Rev. Lett. \textbf{107}, 212501 (2011).

\bibitem{KamLAND}  A.\ Gando \textit{et al.} (The KamLAND-Zen Collaboration),
Phys. Rev. C \textbf{85}, 045504 (2012).

\bibitem{CUPID-0}  O. Azzolini \textit{et al.} (The CUPID-0 Collaboration),
Phys. Rev.\ Lett. \textbf{123}, 262501 (2019).

\bibitem{CUPID-Mo}  E. Armengaud \textit{et al.},
Eur. Phys. J. C \textbf{80}, 674 (2020).

\bibitem{dep2020}F. F. Deppisch, L. Graf, and F. \v{S}imkovic,
Phys. Rev. Lett. \textbf{125}, 171801 (2020).

\bibitem{Chika} Y. Chikashige, R.N. Mohapatra, and R.D.\ Peccei, Phys. Rev.\
Lett. \textbf{45}, 1926 (1980).

\bibitem{gelmini} G.B.\ Gelmini and M. Roncadelli, Phys.\ Lett. B \textbf{99}%
, 411 (1981).

\bibitem{georgi} H.M. Georgi, S.L. Glashow and S. Nussinov, Nucl.\ Phys.\ B 
\textbf{193}, 297 (1981).

\bibitem{CERN} The ALEPH Collaboration, The DELPHI Collaboration, The L3
Collaboration, The OPAL Collaboration, The SLD Collaboration, The LEP
Electroweak Working Group, and The SLD\ Electroweak and Heavy Flavor Groups,
Phys. Rep. \textbf{427}, 257 (2006).

\bibitem{bamert} P. \ Bamert, C. Burgess, and R. Mohapatra, Nucl. Phys. B 
\textbf{449}, 25 (1995).

\bibitem{carone} C.D. Carone, Phys. Lett. B \textbf{308}, 85 (1993).

\bibitem{burgess} C. Burgess and J. Cline, in Proceedings of the First
International Conference on Nonaccelerator Physics, Bangalore, India, 1994,
ed. by R. Cowsik (World Scientific, Singapore, 1995).

\bibitem{mohapatra} R. Mohapatra, A.\ Perez-Lorenzana, and C.D.S. Pires,
Phys.\ Lett. B \textbf{491}, 143 (2000).

\bibitem{kotila-maj} J.\ Kotila, J. Barea and F.\ Iachello, Phys. Rev. C 
\textbf{91}, 064310 (2015).

\bibitem{kotila} J. Kotila and F.\ Iachello, Phys. Rev. C \textbf{85},
034316 (2012).

\bibitem{doi} M. Doi, T. Kotani, and E. Takasugi, Prg. Theor. Phys. Suppl. 
\textbf{83}, 1 (1985).

\bibitem{tomoda} T. Tomoda, Rep. Prog. Phys. \textbf{54}, 53 (1991).

\bibitem{suhonen} J. Suhonen and O. Civitarese, Phys.\ Rep. \textbf{300},
123 (1998).

\bibitem{hirsch} M. \ Hirsch, H. V. Klapdor-Kleingrothaus, S.G. \ Kovalenko
and H. P\"{a}s, Phys. Lett B \textbf{372}, 8 (1996).

\bibitem{NEMO3-1}  R. Arnold \textit{et al.} (The NEMO3 Collaboration), Phys.
Rev. Lett. \textbf{107}, 062504 (2011).

\bibitem{barea1} J.\ Barea and F.\ Iachello, Phys.\ Rev. C \textbf{79},
044301 (2009).

\bibitem{barea2} J.\ Barea, J. Kotila and F.\ Iachello, Phys. Rev. C \textbf{%
87}, 014315 (2013).

\bibitem{barea3} J.\ Barea, J. \ Kotila and F.\ Iachello, C \textbf{91},
034304 (2015).

\bibitem{simkovic} F. \v{S}imkovic, G.\ Pantis, J.D.\ Vergados, and A.
Faessler, Phys. Rev. C \textbf{60}, 055502 (1999).

\bibitem{delloro} S. Dell'Oro, S. Marcocci, and F. Vissani, Phys. Rev. D 
\textbf{90}, 033005 (2014).

\bibitem{CUORE-1}  C. Arnaboldi \textit{et al.} (The CUORE\ Collaboration),
Phys.\ Lett. B \textbf{557}, 167 (2003).

\bibitem{KamLAND-1}  A.\ Gando \textit{et al.} (The KamLAND-Zen Collaboration)%
, Phys. Rev. C \textbf{86}, 021601 (2012).

\bibitem{EXO-1} J.B. Albert \textit{et al.} (The EXO-200 Collaboration),
Phys. Rev. D \textbf{90}, 092004 (2014).

\bibitem{GERDA-1} S.\ Hemmer (GERDA Collaboration), The European Physical\
Journal Plus \textbf{130}, 139 (2015).
\end{thebibliography}
\end{document}